\documentclass[12pt]{article}
\usepackage{fullpage,epsfig,fancyheadings}
\usepackage[psamsfonts]{euscript}

\usepackage{epic}

\usepackage{amstex}

\newcommand{\beq}{\begin{equation}}
\newcommand{\eeq}{\end{equation}}

\newcommand{\K}{{(K^{*})}}

\begin{document}
\vspace*{-.6in}
\thispagestyle{empty}
\begin{flushright}
CALT-68-2018\\
DOE RESEARCH AND\\
DEVELOPMENT REPORT
\end{flushright}
\baselineskip = 20pt

\vspace{.5in}
{\Large
\begin{center}
Chiral Perturbation Theory for $\tau
\rightarrow \rho \pi\nu_\tau$, $\tau \rightarrow K^* \pi \nu_\tau$, and $\tau
\rightarrow \omega \pi \nu_\tau$ \footnote{Work supported in part by the U.S.
Dept. of Energy under Grant No. DE-FG03-92-ER40701.}
\end{center}}
\vspace{.4in}

\begin{center}
Hooman Davoudiasl and Mark B. Wise\\
\emph{California Institute of Technology, Pasadena, CA  91125 USA}
\end{center}
\vspace{1in}

\begin{center}
\textbf{Abstract}
\end{center}
\begin{quotation}
\noindent We use heavy vector meson $SU(2)_L \times SU(2)_R$ chiral
perturbation theory to predict differential decay distributions for $\tau
\rightarrow \rho \pi \nu_\tau$ and $\tau \rightarrow K^* \pi \nu_\tau$ in the
kinematic region where  $p_V \cdot p_\pi/m_V$ (here $V = \rho$ or $K^*$) is
much smaller than the chiral symmetry breaking scale.  Using the large number
of colors limit we also predict the rate for $\tau \rightarrow \omega \pi
\nu_\tau$ in this region (now $V = \omega$).  Comparing our prediction with
experimental data, we determine one of the coupling constants in the heavy
vector meson chiral Lagrangian.
\end{quotation}
\vfil

\newpage

\pagenumbering{arabic} 

\section{Introduction}

Chiral perturbation theory provides a systematic method for describing the
interactions of hadrons at low momentum.  It applies not only to strong
interactions of the pseudo-Goldstone bosons, $\pi, K$, and $\eta$, with
themselves (e.g., $\pi\pi$ scattering), but also to the interactions of the
pseudo-Goldstone bosons with heavy matter fields like nucleons$^{1)}$ and
hadrons containing a heavy charm or bottom quark.$^{2)}$

Recently, chiral perturbation theory has been applied to describe strong
interactions of the lowest lying vector mesons $\rho, K^*, \omega,$ and $\phi$
with the pseudo-Goldstone bosons.$^{3)}$  The vector mesons were treated as
heavy and an effective Lagrangian based on the $SU(3)_L \times SU(3)_R$ chiral
symmetry was given for couplings between the vector mesons and the
pseudo-Goldstone bosons.  At leading order in the derivative expansion, the
chiral Lagrangian has two coupling constants $g_1$ and $g_2$ that are related
in the large $N_c$ (i.e., number of colors) limit.$^{4)}$  While it is known
from the value of the octet singlet mixing angle and the smallness of the $\phi
\rightarrow \rho \pi$ amplitude that the $N_c \rightarrow \infty$ relation,
$g_1 = 2g_2/\sqrt{3}$, is a reasonable approximation, the value of $g_2$ has
not been determined.

In this paper, we use heavy vector meson chiral perturbation theory to study
the
decays $\tau \rightarrow \rho \pi \nu_\tau$ and $\tau \rightarrow K^*
\pi\nu_\tau$ in the kinematic regime where the pion is ``soft'' in the vector
meson's rest frame.  At the present time, there is little experimental
information that  bears on the applicability of chiral perturbation theory for
vector meson interactions.  These $\tau$ decays provide an interesting way to
test whether low orders in the momentum expansion yield a good approximation.
Using the large $N_c$ limit, we also predict the differential decay rate for
$\tau \rightarrow \omega\pi\nu_\tau$, in the kinematic regime where the pion is
soft in the $\omega$ rest frame.  In heavy vector meson chiral perturbation
theory, this decay amplitude is dominated by a rho pole and is proportional to
$g_2^2$.  Comparing with experimental data$^{5)}$, we find that $g_2 \simeq
0.6$.  An important aspect of this work is that we will only
use chiral $SU(2)_L \times SU(2)_R$ symmetry and consequently do not treat the
strange quark mass as small.

The decays $\tau \rightarrow \rho \pi \nu_\tau, \tau \rightarrow K^* \pi
\nu_\tau$, and $\tau \rightarrow \omega\pi\nu_\tau$  result in final hadronic
states that contain three and four pseudo-Goldstone
bosons.  The amplitude for the vector and axial currents to produce
pseudo-Goldstone
bosons is determined by ordinary chiral perturbation theory $^{6)}$ but only in
a limited kinematic region where their invariant mass is small
compared with the chiral symmetry breaking scale.  The situation is similar for
heavy vector meson chiral perturbation theory.  It partially constrains the
multi pseudo-Goldstone boson amplitudes in a small (but different) part of the
available phase space.  This paper is meant to illustrate the usefulness of
heavy vector meson chiral perturbation theory for $\tau$ decay.  Since the
$\rho$ and $K^*$ widths are  not negligible, a more complete calculation that
includes vector meson decay and interference
between different vector meson amplitudes that give the same three
pseudo-Goldstone boson final hadronic state may be necessary for a detailed
comparison with experiment in these cases.

For the $\tau$ decays $\tau \rightarrow \rho \nu_\tau, \tau \rightarrow K^*
\nu_\tau, \tau \rightarrow \rho \pi \nu_\tau$, $\tau \rightarrow K^* \pi
\nu_\tau$, and $\tau \rightarrow \omega\pi\nu_\tau$, we need matrix elements of
the left-handed currents $\bar d
\gamma_\mu (1 - \gamma_5)u$ and $\bar s \gamma_\mu (1 - \gamma_5)u$ between the
vacuum and a vector meson or a vector meson and a low momentum pion.  In the
next section, we derive the hadron level operators that represent these
currents
in chiral perturbation theory.  Section 3 contains expressions for the $\tau
\rightarrow \rho \pi \nu_\tau$, $\tau \rightarrow K^* \pi \nu_\tau$, and $\tau
\rightarrow \omega \pi \nu_\tau$ differential decay rates.  Concluding remarks
are made in Section 4.

\section{Chiral Perturbation Theory For Vector Mesons}

An effective Lagrangian based on $SU(2)_L \times SU(2)_R$ chiral symmetry that
describes the interactions of $\rho$ and $K^*$ vector mesons with pions can be
derived in the standard way.  The pions are incorporated into a $2 \times 2$
special unitary matrix
\begin{equation}
\Sigma = \exp (2i \Pi/f),
\end{equation}
where
\begin{equation}
\Pi = \left[ \begin{array}{cc}
\pi^0/\sqrt{2} & \pi^+\\
\pi^- & - \pi^0/\sqrt{2}
\end{array} \right] .
\end{equation}
Under chiral $SU(2)_L \times SU(2)_R, \Sigma \rightarrow L\Sigma R^\dagger$,
where $L\in SU(2)_L$ and $R\in SU(2)_R$.  At leading order in chiral
perturbation theory, $f$ can be identified with the pion decay constant $f_\pi
\simeq 132 MeV$.  For describing the interactions of the pions with other
fields it is convenient to introduce
\begin{equation}
\xi = \exp \left({i\Pi\over f}\right) = \sqrt{\Sigma}.
\end{equation}
Under chiral $SU(2)_L \times SU(2)_R$,
\begin{equation}
\xi \rightarrow L \xi U^\dagger = U \xi R^\dagger,
\end{equation}
where $U$ is a complicated function of $L,R$, and the pion fields $\Pi$.
However, in the special case of transformations where $L = R = V$ in the
unbroken $SU(2)_V$ vector subgroup, $U = V$.

The $\rho$ fields are introduced as a $2 \times 2$ matrix
\begin{equation}
R_\mu = \left[ \begin{array}{cc}
\rho^0_\mu/\sqrt{2} & \rho^+_\mu\\
\rho^-_\mu & - \rho^0_\mu/\sqrt{2}
\end{array} \right] ,
\end{equation}
and the $K^*, \bar K^*$ fields as doublets
\begin{equation}
K^*_\mu = \left[ \begin{array}{c}
K^{*+}_\mu\\
K^{*0}_\mu
\end{array} \right], \qquad \bar K_\mu^* = \left[ \begin{array}{c}
K^{*-}_\mu\\
\bar K^{*0}_\mu
\end{array} \right] .
\end{equation}
Under chiral $SU(2)_L \times SU(2)_R$,
\begin{equation}
R_\mu \rightarrow U R_\mu U^\dagger, \quad K_\mu^* \rightarrow U K_\mu^*,~~
\bar K_\mu^* \rightarrow U^* \bar K^*_\mu .
\end{equation}
The doublets $K^*_\mu$ and $\bar K_\mu^*$ are related by charge conjugation
which acts on the fields as follows:
\begin{equation}
CR_\mu C^{-1} = - R_\mu^T,\quad C K_\mu^* C^{-1} = - \bar K_\mu^*,\quad C\xi
C^{-1} = \xi^T.
\end{equation}

We construct an effective Lagrangian for strong transitions of the form $V
\rightarrow V' X$, where $V$ and $V'$ are vector mesons and $X$ is either the
vacuum or one or more soft pions.  The vector meson fields are treated as heavy
with fixed four velocity $v^\mu, v^2 = 1$, satisfying the constraint $v \cdot R
= v \cdot K^* = v \cdot \bar K^* = 0$.  The chiral Lagrange density has the
general structure
\begin{equation}
{\cal L} = {\cal L}_{kin} + {\cal L}_{int} + {\cal L}_{mass} - {i\over 2} {\cal
L}_{width} .
\end{equation}
The interaction terms are
\begin{equation}
{\cal L}_{int} = ig_2^{(\rho)} Tr (\{R_\mu^\dagger, R_\nu\}
A_\lambda)v_\sigma \epsilon^{\mu\nu\lambda\sigma}
+ ig_2^{(K^*)} \bar K_\mu^{*\dagger} A_\lambda^T \bar K_\nu^* v_\sigma
\epsilon^{\mu\nu\lambda\sigma}
+ig_2^{(K^*)} K_\mu^{*\dagger} A_\lambda K_\nu^* v_\sigma
\epsilon^{\mu\nu\lambda\sigma},
\end{equation}
where
\begin{equation}
A_\lambda = {i\over 2} (\xi\partial_\lambda \xi^\dagger - \xi^\dagger
\partial_\lambda \xi).
\end{equation}
Comparing with the Lagrange density in eq. (11) of Ref. [3], we find that in
the
case of $SU(3)_L \times SU(3)_R$ symmetry $g_2^{(\rho)} = g_2^{(K^{*})} = g_2$,
at leading order in $SU(3)_L \times SU(3)_R$ chiral perturbation theory.  At
higher orders, integrating out the kaons will lead to a difference between
$g_2^{(\rho)}$ and $g_2^{(K^{*})}$.  Note that for the vector mesons
$\rho_\mu^{-\dagger} \not= \rho_\mu^+$, etc.  In heavy vector meson chiral
perturbation theory, $\rho_\mu^+$ destroys a $\rho^+$, but it does not create
the corresponding antiparticle.  The field $\rho_\mu^{-\dagger}$ creates a
$\rho^-$.

The kinetic terms are
\begin{eqnarray}
{\cal L}_{kin}  = &-& i Tr R_\mu^\dagger v \cdot \partial R^\mu -i Tr
R_\mu^\dagger [v \cdot V, R^\mu]
-i K_\mu^{*\dagger} v \cdot \partial K^{*\mu} -i K_\mu^{*\dagger} v \cdot V
K^{*\mu} \nonumber\\
&-& i \bar K_\mu^{*\dagger} v \cdot \partial \bar K^{*\mu} + i \bar
K_\mu^{*\dagger} v \cdot V^T \bar K^{*\mu},
\end{eqnarray}
where
\begin{equation}
V_\nu = {1\over 2} (\xi \partial_\nu \xi^\dagger + \xi^\dagger \partial_\nu
\xi).
\end{equation}
The mass terms are
\begin{eqnarray}
{\cal L}_{mass} &=&\lambda_2^{(\rho)} Tr (\{R_\mu^\dagger, R^\mu\}
M_\xi)
+\lambda_2^{(K^{*})} K_\mu^{*\dagger} M_\xi K^\mu
+\lambda_2^{(K^{*})} \bar K_\mu^{*\dagger} M_\xi^T \bar
K^{*\mu}\nonumber\\
&+&\sigma_8^{(\rho)} Tr (M_\xi) Tr (R_\mu^\dagger R^\mu)
+\sigma_8^{(K^{*})} Tr (M_\xi) K_\mu^{*\dagger} K^\mu
+\sigma_8^{(K^{*})} Tr (M_\xi) \bar K_\mu^{*\dagger} \bar K^{*\mu}.
\end{eqnarray}
In eq. (14)
\begin{equation}
M_\xi = {1\over 2} (\xi M\xi + \xi^\dagger M \xi^\dagger),
\end{equation}
where $M = diag (m_u, m_d)$ is the $2\times 2$ quark mass matrix.  At leading
order in $SU(3)_L \times SU(3)_R$ chiral perturbation theory, the couplings in
eq. (14) are related to those in Ref. [3] by
\begin{equation}
\lambda_2^{(\rho)} = \lambda_2^{(K^{*})} = \lambda_2 ~{\rm and}~
\sigma_8^{(\rho)} = \sigma_8^{(K^{*})} = \sigma_8.
\end{equation}

The $\rho$ and $K^*$ are not stable.  In heavy vector meson chiral perturbation
theory, their widths appear as antihermitian terms in the Lagrange density (9).
Since the $\rho$ and $K^*$ widths vanish in the large $N_c$ (i.e., number of
colors) limit and are comparable with the pion mass, we treat the widths as of
order one derivative (the mass terms in (14) go like two derivatives and are
less important in chiral perturbation theory than the terms in ${\cal L}_{kin},
{\cal L}_{int}$ and ${\cal L}_{width}$).  The width terms are
\begin{equation}
{\cal L}_{width} =\Gamma^{(\rho)} Tr R_\mu^\dagger R^\mu + \Gamma^{(K^{*})}
K_\mu^{*\dagger} K^{*\mu}
+\Gamma^{(K^{*})} \bar K_\mu^{*\dagger} \bar K^{*\mu}.
\end{equation}
In the $SU(3)$ limit $\Gamma^{(\rho)} = \Gamma^{(K^{*})}$, however, the
physical values of the widths $\Gamma^{(\rho)} = 151 MeV$ and $\Gamma^{(K^{*})}
= 50 MeV$ are far from this situation.  In heavy vector meson chiral
perturbation theory, the vector meson propagator is
\begin{equation}
{-i (g^{\mu\nu} - v^\mu v^\nu)\over v \cdot k + i\Gamma/2},
\end{equation}
where $\Gamma$ is the corresponding width.  Note that we are treating the
vector meson widths differently than Ref. [3].  In Ref. [3], chiral $SU(3)_L
\times SU(3)_R$ was used and since the vector meson widths are small compared
with the kaon mass they were treated as of the order of a light quark mass or,
equivalently, two derivatives.  Hence, in Ref. [3], the widths could be
neglected in the propagator at leading order in chiral perturbation theory.

At the quark level, the effective Hamiltonian density for weak semileptonic
$\tau$ decay is
\begin{equation}
H_W = {G_F\over\sqrt{2}} V_{ud} \bar\nu_\tau \gamma_\mu (1 - \gamma_5) \tau
\bar d \gamma^\mu (1 - \gamma_5)u
+ {G_F\over\sqrt{2}} V_{us} \bar\nu_\tau \gamma_\mu (1 - \gamma_5)\tau \bar s
\gamma^\mu (1 - \gamma_5)u,
\end{equation}
where $G_F$ is the Fermi constant and $V_{ud}$ and $V_{us}$ are elements of the
Cabibbo-Kobayashi-Maskawa matrix where, experimentally, $|V_{ud}| \simeq 1$ and
$|V_{us}| \simeq 0.22.$  At leading order in chiral perturbation theory, we
need
to represent the currents $\bar d \gamma^\mu (1 - \gamma_5)u$ and $\bar s
\gamma^\mu (1 - \gamma_5)u$ by operators involving the hadron fields that
transform respectively as $(3_L, 1_R)$ and $(2_L, 1_R)$ under chiral $SU(2)_L
\times SU(2)_R$ and contain the least number of derivatives or insertions of
the light quark mass matrix.  These operators are
\renewcommand{\theequation}{20a}
\begin{equation}
\bar d \gamma_\mu (1 - \gamma_5) u = {f_\rho\over\sqrt{2m_\rho}} Tr
R_\mu^\dagger \xi^\dagger \left(\begin{array}{cc} 0 & 0\\ 1 & 0 \end{array}
\right) \xi,
\end{equation}
and
\renewcommand{\theequation}{20b}
\begin{equation}
\bar s \gamma_\mu (1 - \gamma_5 )u = {f_{K^{*}}\over \sqrt{2m_{K^{*}}}} \bar
K_\mu^{*\dagger} \xi^T
\left(\begin{array}{c} 1\\ 0 \end{array} \right).
\end{equation}
\renewcommand{\theequation}{\arabic{equation}}
\setcounter{equation}{20}
The coefficients are fixed in terms of the vector meson decay constants
$f_\rho$ and $f_{K^{*}}$ by the matrix elements $<K^{*-} |\bar s \gamma_\mu (1
- \gamma_5) u|0>$ and $<\rho^-  | \bar d \gamma_\mu (1 - \gamma_5) u|0>$, which
are equal to $f_{K^{*}} \epsilon_\mu^{*}$ and $f_\rho \epsilon_\mu^{*}$
respectively,
and follow from equations (20a) and (20b) by setting $\xi$ equal to unity.
(Note
that because of the parity invariance of the strong interactions the axial
currents do not contribute to these matrix elements.)

In the large $N_c$ limit, couplings involving the $\omega$ are related to those
involving the $\rho$.
They can be derived from the Lagrange densities in eqs. (10), (12), (14) and
the expression for the current in eq. (20a) by replacing the isospin triplet
matrix $R_\mu$ by the quartet matrix
\begin{equation}
Q_\mu =  \left[\begin{array}{cc}
\rho^0_\mu/\sqrt{2} + \omega_\mu/\sqrt{2} & \rho_\mu^+\\
\rho_\mu^- & -\rho_\mu^0/\sqrt{2} + \omega_\mu/\sqrt{2} \end{array} \right].
\end{equation}
However, the effect of the $\omega$ width cannot be included by replacing
$R_\mu$ in eq. (17) with $Q_\mu$.  Since the widths vanish in the large $N_c$
limit, a
separate term $\Gamma^{(\omega)} \omega_\mu^\dagger \omega^\mu$ must be added
to eq.
(17).  Experimentally, $\Gamma^{(\omega)} = 8.4 MeV$.

\section{Differential Decay Rates}

The amplitude for $\tau \rightarrow \rho \pi  \nu_\tau$ follows from the
Feynman diagram for the vacuum to $\rho\pi$ matrix element of the current
shown in Fig. 1.  Note that there is no pole diagram since the Lagrange density
(10) has no $\rho\rho\pi$ coupling.  The invariant matrix element is
\begin{equation}
{\cal M}(\tau \rightarrow \rho^0 \pi^- \nu_\tau) = {G_F V_{ud} f_\rho\over
f_\pi} \bar u_\nu \gamma^\mu \epsilon^*_\mu (\rho) (1-\gamma_5) u_\tau,
\end{equation}
where $u_{\nu,\tau}$ are four component spinors for the neutrino and tau.

It is convenient to express the differential decay distribution in terms of the
$\rho \pi$ mass $s = (p_\rho + p_\pi)^2$ and the angle $\theta$ between the
$\rho$ direction and the $\tau$ direction in the $\rho-\pi$ center of mass
frame.  Then the differential decay rate is
\[
{d\Gamma (\tau\rightarrow\rho^0 \pi^- \nu_\tau)\over ds d\cos\theta} = {G_F^2
|V_{ud}|^2 f_\rho^2 m_\tau\over 2^7 f_\pi^2 \pi^3}
\left({1 - {s\over  m_{\tau}^{2}}}\right)
\sqrt{{(s - m_\rho^2 + m_\pi^2)^2 - 4m_\pi^2 s\over 4s^2}}\]
\begin{equation}
\times[A(s) + B(s) \cos\theta + C(s) \cos^2\theta],
\end{equation}
where the dimensionless functions $A(s), B(s)$ and $C(s)$ are
\renewcommand{\theequation}{24a}
\begin{equation}
A(s) = {1\over 8s^2 m_\rho^2}\left({1 - {s\over  m_{\tau}^{2}}}\right)
[(s+m_\rho^2 - m_\pi^2)^2 (s+m_\tau^2) + 4s^2 m_\rho^2],
\end{equation}
\renewcommand{\theequation}{24b}
\begin{equation}
B(s) = - {m_\tau^2\over 4s^2 m_\rho^2}
\left({1 - {s\over  m_{\tau}^{2}}}\right)
(s+m_\rho^2 - m_\pi^2) \sqrt{(s-m_\rho^2 + m_\pi^2)^2 - 4m_\pi^2 s},
\end{equation}
\renewcommand{\theequation}{24c}
and
\begin{equation}
C(s) = {m_\tau^2\over 8s^2 m_\rho^2}\left({1 - {s\over  m_{\tau}^{2}}}\right)^2
[(s-m_\rho^2 + m_\pi^2)^2 - 4m_\pi^2 s].
\end{equation}
\renewcommand{\theequation}{\arabic{equation}}
\setcounter{equation}{24}
The differential decay rate is the same for the $\rho^- \pi^0$ mode.  Our
expression for the invariant matrix element in eq. (22) was derived using heavy
vector meson chiral perturbation theory, which is an expansion in
$m_\pi/m_\rho$ and $v \cdot p_\pi/m_\rho$.  In $A$, $B$, and $C$, terms
suppressed by
powers of these quantities should be neglected.  To focus on the kinematic
region where chiral perturbation theory is valid, it is convenient to change
from the variable $s$ to the dimensionless variable $x = v \cdot p_\pi/m_\pi$,
using,
\begin{equation}
s = m_\rho^2 + m_\pi^2 + 2m_\pi m_\rho x.
\end{equation}
Then expanding in $(m_\pi/m_\rho)$,
\renewcommand{\theequation}{26a}
\begin{equation}
A(x) \simeq {1\over 2} \left(1 - {m_\rho^2\over m_\tau^2}\right) \left[2 +
{m_\tau^2\over m_\rho^2}\right],
\end{equation}
\renewcommand{\theequation}{26b}
\begin{equation}
B(x) \simeq - \left({m_\pi\over m_\rho}\right) \left({m_\tau^2\over m_\rho^2}
-1\right) \sqrt{x^2 - 1},
\end{equation}
and
\renewcommand{\theequation}{26c}
\begin{equation}
C(x) \simeq {1\over 2} \left({m_\pi^2\over m_\rho^2}\right)
\left({m_\tau^2\over m_\rho^2}\right) \left(1 - {m_\rho^2\over
m_\tau^2}\right)^2 (x^2 - 1).
\end{equation}
\renewcommand{\theequation}{\arabic{equation}}
\setcounter{equation}{26}
Hence, $B$ and $C$ are negligible compared with $A$ and our expression for the
differential decay rate becomes
\begin{equation}
{d\Gamma(\tau \rightarrow \rho^0 \pi^- \nu_\tau)\over dx d\cos\theta} = {G_F^2
|V_{ud}|^2 f_\rho^2 m_\tau m_\pi^2\over 2^7 f_\pi^2 \pi^3} \sqrt{x^2 - 1}
\left(1 - {m_\rho^2\over m_\tau^2}\right)^2 \left[2+ {m_\tau^2\over
m_\rho^2}\right].
\end{equation}
Normalizing to the $\tau \rightarrow \rho^- \nu_\tau$ width gives the simple
expression
\begin{equation}
{1\over \Gamma (\tau \rightarrow \rho^- \nu_\tau)} {d\Gamma (\tau \rightarrow
\rho^0 \pi^- \nu_\tau)\over dx} = \left({m_\pi\over f_\pi} \right)^2 {\sqrt{x^2
-
1}\over 4\pi^2}.
\end{equation}
It seems reasonable that lowest order chiral perturbation theory will be a
useful approximation in the region $x\in [1,2]$.  Integrating $x$ over this
region gives a $\tau\rightarrow\rho^0 \pi^- \nu_\tau$ width that is 0.03 times
the $\tau\rightarrow\rho^- \nu_\tau$ width.

The amplitude for $\tau \rightarrow K^* \pi\nu_\tau$ follows from the Feynman
diagrams for the vacuum to $K^*\pi$ matrix element of the left-handed current
shown in Fig. 2.  In this case, there is a  pole contribution proportional to
the $K^* K^*\pi$ coupling $g_2^{(K^{*})}$.  The resulting invariant matrix
element is
\[
{\cal M} (\tau \rightarrow \bar K^{*0} \pi^- \nu_\tau) = {G_F V_{us}
f_{K^{*}}\over\sqrt{2} f_\pi} \bar u_\nu \gamma_\mu (1 - \gamma_5) u_\tau
\]
\begin{equation}
\times \left[\epsilon^{*\mu} (K^*) + {ig_2^{(K^{*})}
\epsilon^{\nu\mu\beta\sigma}\over (v \cdot p_\pi + i\Gamma^{(K^{*})}/2)}
p_{\pi\beta} v_\sigma \epsilon_\nu^* (K^*)\right].
\end{equation}
The term proportional to $g_2^\K$ arises from the pole diagram and it
corresponds to the $p$-wave $K^*\pi$ amplitude.  In the nonrelativistic
constituent quark model$^{7)}$ $g_2^\K = 1$.
Following the same procedure as for the $\tau \rightarrow \rho \pi \nu_\tau$
case, we arrive at the differential decay rate
\newpage
\[
{d\Gamma (\tau \rightarrow \bar K^{*0} \pi^- \nu_\tau)\over dx d\cos\theta} =
{G_F^2 |V_{us}|^2 f_{K^{*}}^2 m_\pi^2 m_\tau\over 2^8 f_\pi^2 \pi^3} \sqrt{x^2
- 1} \left(1 - {m_{K^{*}}^2\over m_\tau^2}\right)^2\]
\[ \times \left\{ \left({m_\tau^2\over m_{K^{*}}^2} +2\right) +
{g_2^{(K^{*})2}\over x^2 (1 + \gamma^2)} \left({m_\tau^2\over m_{K^{*}}^2}
+1\right) (x^2 - 1) \right.\]
\begin{equation}
\left. + {4g_2^{(K^{*})}\over x(1 +\gamma^2)} \sqrt{x^2 - 1} \cos\theta -
{g_2^{(K^{*})^2}\over x^2 (1 + \gamma^2)} (x^2 - 1) \left({m_\tau^2\over
m_{K^{*}}^2} -1 \right) \cos^2 \theta \right\}.
\end{equation}
In eq. (30)
\begin{equation}
\gamma = \Gamma^{(K^{*})} /(2xm_\pi).
\end{equation}
In this case, $s = m_{K^{*}}^2 + m_\pi^2 + 2m_\pi m_{K^{*}} x$.  The rate for
$\tau\rightarrow K^{*-} \pi^0 \nu_\tau$ is one half the rate for
$\tau\rightarrow \bar K^{*0} \pi^- \nu_\tau$.

Normalizing to the $\tau \rightarrow K^{*-} \nu_\tau$ decay width and
integrating
over $x\in$  [1,2], eq. (30) gives
\[
{1\over\Gamma(\tau\rightarrow K^{*-} \nu_\tau)} \int_1^2 dx
{d\Gamma(\tau\rightarrow \bar K^{*0} \pi^- \nu_\tau)\over dx d\cos\theta}
\simeq 7.5 \times 10^{-3} [(1 + 0.48 g_2^{(K^{*})^2})\]
\begin{equation}
+ 0.51 g_2^{(K^{*})} \cos\theta - 0.28 g_2^{(K^{*})^2} \cos^2 \theta].
\end{equation}
The shape of the $\bar K^{*0} \pi^- \nu_\tau$ decay distribution in
$\cos\theta$ depends on the value of $g_2^{(K^{*})}$ and it may be possible at
a tau-charm or $B$ factory to determine this coupling from a study of $\tau
\rightarrow K^* \pi \nu_\tau$ decay.  In the $SU(3)$ limit
$g_2^{(K^{*})}=g_2^{(\rho)}$ and in what follows we discuss how to determine
$g_2^{(\rho)}$ in the large $N_c$ limit.

Using both heavy vector meson chiral perturbation theory and the large $N_c$
limit, the amplitude for $\tau \rightarrow \omega\pi\nu_\tau$ follows from the
Feynman diagram for vacuum to $\omega\pi$ matrix element of the left handed
current in Fig. 3.  In this case, there is only a pole graph and the invariant
matrix element is
\begin{equation}
{\cal M}(\tau \rightarrow \omega\pi^-\nu_\tau) =  {G_FV_{ud} f_\rho\over f_\pi}
\bar
u_\nu \gamma_\mu (1 - \gamma_5)u_\tau  \left[{ig_2^{(\rho)}\over (v \cdot p_\pi
+ i \Gamma^{(\rho)}/2)} \quad
\epsilon^{\nu\mu\beta\sigma} p_{\pi\beta} v_\sigma \epsilon_\nu^*
(\omega)\right].
\end{equation}
Here, the difference between the $\rho$ and $\omega$ masses is neglected as is
appropriate in the large $N_c$ limit.  The resulting differential decay rate is
\newpage
\[
{d\Gamma(\tau \rightarrow \omega\pi^-\nu_\tau)\over dx d\cos\theta} =
{G_F^2|V_{ud}|^2 f_\rho^2 m_\pi^2 m_\tau\over 2^7 f_\pi^2 \pi^3} (x^2 -
1)^{3/2} \left(1-{m_\omega^2\over m_\tau^2}\right)^2\]
\begin{equation}
\times {g_2^{(\rho)^{2}}\over x^2 (1 + \gamma^2)} \left[\left({m_\tau^2\over
m_\omega^2} +1\right) - \left({m_\tau^2\over m_\omega^2} -1\right) \cos^2
\theta\right]
\end{equation}
where now
\begin{equation}
\gamma = \Gamma^{(\rho)}/(2xm_\pi),
\end{equation}
and $s = m_\omega^2 + m_\pi^2 + 2m_\omega m_\pi x$.  Integrating over
$\cos\theta$ and dividing by the rate for $\tau \rightarrow \rho \nu_\tau$
give (again we neglect the difference between the $\rho$ and $\omega$ masses)
the simple expression,
\begin{equation}
{1\over\Gamma(\tau \rightarrow \rho^- \nu_\tau)} {d\Gamma (\tau \rightarrow
\omega \pi^- \nu_\tau)\over dx} = \left({m_\pi\over f_\pi}\right)^2
{(x^2-1)^{3/2} g_2^{(\rho)^2}\over 6\pi^2 x^2 (1+\gamma^2)}.
\end{equation}
Ref. [5] plots the differential decay rate as a function of the $\omega\pi$
invariant mass (see Fig. (3b)).  The first bin corresponds to $x \leq 1.7$.
Integrating the $\tau \rightarrow \omega \pi \nu_\tau$ differential decay rate
over $x\in$[1, 1.7] and comparing with the experimental rate in this
region$^{8)}$ give $|g_2^{(\rho)}| \simeq 0.57$.  If both the first and second
bins are included the region corresponds to $x \in$[1, 2.7] and
integrating over this region gives $|g_2^{(\rho)}| \simeq 0.65$.  It is not
likely that lowest order chiral perturbation theory will be a good
approximation for values of $x$ greater than this.

\section{Concluding Remarks}

In this paper, we have studied the decay modes $\tau \rightarrow \rho \pi
\nu_\tau$, $\tau \rightarrow K^* \pi \nu_\tau$, and $\tau \rightarrow
\omega\pi \nu_\tau$, using heavy vector meson
chiral perturbation theory.  Eqs. (27), (30) and (34) are our main results.
Our predictions are valid in the kinematic region where the pion is soft in the
vector meson rest frame.  For these modes, vector meson decay results in three
or four pseudo-Goldstone boson hadronic final states, and heavy vector meson
chiral perturbation theory restricts these amplitudes in a small part of phase
space.  This is similar to applications of ordinary chiral perturbation theory
which are valid in a different small kinematic region.

Modes similar to those discussed in this paper, such as $\tau \rightarrow \rho
K \nu_\tau$, can also be studied, using chiral perturbation theory.  They will
be related to those we considered in chiral $SU(3)_L \times SU(3)_R$.  Using
chiral $SU(3)_L \times SU(3)_R$, the left handed current $J_{L\lambda}^A = \bar
q T^A \gamma_\lambda (1 - \gamma_5) q$ is represented by
\begin{equation}
J_{L\lambda}^A = {f_V\over\sqrt{2m_V}} Tr (O_\lambda^\dagger \xi^\dagger T^A
\xi),
\end{equation}
where $O_\lambda$ is the $3\times 3$ octet matrix of vector meson fields.

We found a branching ratio for $\tau \rightarrow \rho^0 \pi^- \nu_\tau$ in the
region where the hadronic mass satisfies $m_{\rho\pi}< 1022 MeV$, of 0.69\%,
and
a branching ratio for $\tau \rightarrow \bar K^{*0}\pi^- \nu_\tau$ in the
region $m_{K^{*}\pi} < 1151 MeV$, of $(0.02 + 0.008 g_2^{(K^{*})^2}) \%$.  It
may be possible to study $\tau \rightarrow K^* \pi \nu_\tau$ decay in the
kinematic region where chiral perturbation theory is valid at a
$\tau$-charm or $B$ factory.$^{9)}$

In $\tau$ decay, the  $\rho\pi$ final hadronic states get a significant
contribution from the $a_1(1260)$ resonance which has a large width of around
400 MeV, while $K^* \pi$ final states get contributions from the $K_1(1270),~
K_1(1400)$, and $K^* (1410)$ which have widths of 90 MeV, 174 MeV, and 227 MeV,
respectively.   Since in our formulation of chiral perturbation theory these
heavier resonances are integrated out, one can take the view that the ``tails''
of their contributions are constrained by our results.   Note that the
$K_1$(1270) has a branching ratio of only 16\% to $K^* \pi$.

The narrow width of the $\omega$ makes $\tau \rightarrow \omega\pi\nu_\tau$
easier to study experimentally than $\tau \rightarrow \rho\pi\nu_\tau$.  Using
both heavy vector meson chiral perturbation theory and the large $N_c$ limit,
we
predicted the differential decay rate for $\tau \rightarrow \omega\pi\nu_\tau$
in the kinematic region where the pion is soft in the $\omega$ rest frame.
Comparing with experimental data, we find that the $\rho \omega\pi$ coupling,
$|g_2^{(\rho)}| \simeq 0.6$.  $\tau \rightarrow \omega \pi \nu_\tau$ decay
proceeds via the vector part of the weak current and the rate for this decay is
related by isospin to the $e^+ e^- \rightarrow \omega\pi^0$ cross section.
Experimental data $^{10)}$ on $e^+ e^-  \rightarrow \omega\pi^0$ lead to a
comparable value for $g_2^{(\rho)}$.

Our predictions for $\tau$ decay amplitudes get corrections suppressed by just
$\sim v \cdot p_\pi/(1GeV)$ from operators with one derivative (e.g., $Tr
O_\lambda^\dagger v \cdot A \xi^\dagger T^A \xi$) that occur in the left-handed
current.  This is different from pseudo-Goldstone boson self interactions where
corrections to leading order results are suppressed by $p^2/(1GeV^2)$, where
$p$ is a typical momentum.  Hence, even in the region, $1< v \cdot p_\pi/m_\pi
<2$, we expect sizeable corrections to our results.  This is particularly true
for the $\rho \pi$ case where this region overlaps with a significant part of
the $a_1$ Breit--Wigner distribution.

We have applied heavy vector meson chiral perturbation theory to $\tau$ decay
and used data on $\tau \rightarrow \omega\pi \nu_\tau$ to determine the
magnitude of the
coupling $g_2$ in the chiral Lagrangian.  The value we extract, $|g_2|
\simeq 0.6$, is not too far from the prediction, $g_2 = 0.75$, of the chiral
quark model.$^{11)}$  The value of $g_2$ is relevant for other processes of
experimental interest.  For example, heavy vector meson chiral perturbation
theory can be used to predict differential decay rates for $D \rightarrow K^*
\pi e^+ \nu_e$ in the kinematic region where both $p_D \cdot p_\pi/m_D$ and
$p_{K^*} \cdot p_\pi/m_{K^{*}}$ are small compared with the chiral symmetry
breaking scale.  Here, one combines chiral perturbation theory for hadrons
containing a heavy quark$^{2)}$ with heavy vector meson chiral perturbation
theory.

\section*{Acknowledgements}

We thank A. Weinstein, J. Urheim, A. Manohar, and A. Leibovich for useful
discussions.
\newpage
\centerline{{\bf Figure Captions}}
\begin{enumerate}
\item  Feynman diagram representing the matrix element of the left-handed
current from the vacuum to $\rho\pi$.  In this case, only the axial current
contributes.

\item  Feynman diagrams representing the matrix element of the left-handed
current from the vacuum to $K^* \pi$.  For the first diagram, the axial current
contributes while for the second pole diagram, the vector current contributes.

\item  Feynman diagram representing the matrix element of the left-handed
current from the vacuum to $\omega\pi$.  In this case, only the vector current
contributes.
\end{enumerate}

\newpage
\input FEYNMAN

\begin{picture}(1000,2000)(0,5000)
\thicklines
\put(11000,-5000){\framebox(400,400)}
\put(11050,-4950){\framebox(300,300)}
\put(11100,-4900){\framebox(200,200)}
\put(11150,-4850){\framebox(100,100)}
\put(11200,-4820){\line(1,0){10000}}
\drawline\scalar[\NE\REG](11200,-4900)[5]

\put(16000,-10500) {\makebox(0,0)[t]{\Large Figure 1}}
\put(22000,-3820) {\makebox(0,0)[t]{\Large$\rho$}}
\put(20500,2500) {\makebox(0,0)[t]{\Large$\pi$}}

\put(0,-26000){\framebox(400,400)}
\put(50,-25950){\framebox(300,300)}
\put(100,-25900){\framebox(200,200)}
\put(150,-25850){\framebox(100,100)}
\put(200,-25820){\line(1,0){10000}}
\drawline\scalar[\NE\REG](200,-25820)[5]

\put(11000,-24000) {\makebox(0,0)[t]{\Large$K^*$}}
\put(9500,-18500) {\makebox(0,0)[t]{\Large$\pi$}}

\put(20000,-26000){\framebox(400,400)}
\put(20050,-25950){\framebox(300,300)}
\put(20100,-25900){\framebox(200,200)}
\put(20150,-25850){\framebox(100,100)}
\put(20180,-25820){\framebox(50,50)}
\put(20200,-25800){\line(1,0){10000}}
\put(24200,-25800){\circle*{500}}
\drawline\scalar[\NE\REG](24200,-25800)[4]

\put(16000,-31480) {\makebox(0,0)[t]{\Large Figure 2}}

\put(22500,-24000) {\makebox(0,0)[t]{\Large$K^*$}}
\put(31000,-20000) {\makebox(0,0)[t]{\Large$\pi$}}
\put(32000,-24000) {\makebox(0,0)[t]{\Large$K^*$}}

\put(11000,-47000){\framebox(400,400)}
\put(11050,-46950){\framebox(300,300)}
\put(11100,-46900){\framebox(200,200)}
\put(11150,-46850){\framebox(100,100)}
\put(11180,-46820){\framebox(50,50)}
\put(11200,-46800){\line(1,0){10000}}
\put(15200,-46800){\circle*{500}}
\drawline\scalar[\NE\REG](15200,-46800)[4]

\put(16000,-52480) {\makebox(0,0)[t]{\Large Figure 3}}

\put(13000,-45000) {\makebox(0,0)[t]{\Large$\rho$}}
\put(22000,-41000) {\makebox(0,0)[t]{\Large$\pi$}}
\put(21200,-45000) {\makebox(0,0)[t]{\Large$\omega$}}

\end{picture}

\newpage
\section*{References}
\begin{enumerate}

\item  P. Langacker and H. Pagels, Phys. Rev. {\bf D10} 2904 (1974);  J.
Bijnens, H. Sonoda, and M.B. Wise, Nucl. Phys. {\bf B261} 185 (1985); J.
Gasser, M.E. Sainio and A. Svarc, Nucl. Phys. {\bf B307}, 779 (1988);
E. Jenkins and A.V. Manohar, Phys. Lett. {\bf B255}, 558 (1991); {\bf 259} 353
(1991).

\item M.B. Wise, Phys. Rev. {\bf D45} 2188 (1992); G. Burdman and J.F.
Donoghue, Phys. Lett. {\bf B280} 287 (1992); T.-M. Yan, et.al., Phys. Rev. {\bf
D46} 1148 (1992); P. Cho, Nucl. Phys. {\bf B396} 183 (1993).

\item E. Jenkins, A.V. Manohar and M.B. Wise, Phys. Rev. Lett. {\bf 75} 2272
(1995).

\item G. 't Hooft, Nucl. Phys. {\bf B72} 461 (1974).

\item R. Balest, et al., (CLEO Collaboration) CLNS 95/1347 (1995) unpublished.

\item R. Decker, M. Finkemeier and E. Mirkes, Phys. Rev. {\bf D50} 6863 (1994).

\item See, for example, J.J.J. Kokkedee, {\it The Quark Model} (Benjamin, New
York 1969).  In the same approximation, the pion nucleon coupling $g_A = 5/3$.
Experimentally, $g_A \simeq 1.25.$

\item In Fig. (3b) of Ref. (4) the function $v$ is plotted.  Using eq. (34) our
prediction for the spectral function $v$ as a function of, $x =
(m_{\omega\pi}^2 - m_\omega^2 - m_\pi^2)/2m_\pi m_\omega$, is
\[v (x) = {1\over 6\pi} \left({m_\pi\over m_\omega}\right) \left({f_\rho\over
f_\pi}\right)^2 {1\over m_\omega^2} {(x^2 - 1)^{3/2}\over x^2 (1 + \gamma^2)}
g_2^{(\rho)^{2}}\]
Experimentally, the $\rho$ decay constant is $f_\rho \simeq (407 MeV)^2$.  In
Fig. (3b), in the first bin, $0.9 GeV < m_{\omega \pi} < 1.0 GeV$, $v = (0.0029
\pm 0.0004)$, and in the second bin $1 GeV < m_{\omega \pi} < 1.1GeV, v =
(0.0156 \pm 0.0018)$.  These errors are only statistical.

\item For a recent discussion of data on $K\pi\pi$ hadronic final states see
J.G. Smith (CLEO Collaboration) Nucl. Phys. B (Proc. Suppl.) {\bf 40} 351
(1995); M. Battle et.al. (CLEO Collaboration), Phys. Rev. Lett. {\bf 73} 1079
(1994).
There, it is reported that $Br(\tau \rightarrow \bar K^0 \pi^- \pi^0 \nu_\tau)
= (0.39 \pm 0.06\pm 0.06)\%; Br(\tau \rightarrow  K^- \pi^0 \pi^0 \nu_\tau) =
(0.14 \pm 0.1\pm 0.03)\%$.

\item S.I. Eidelman and V.N. Ivanchenko, Nucl. Phys. B (Proc. Suppl.) {\bf 40}
131 (1995).

\item A.V. Manohar and H. Georgi, Nucl. Phys. {\bf B234} 189 (1984).
\end{enumerate}

\end{document}